\documentclass[doublecol,linenumbers]{epl2} 
\usepackage{amsmath}

\title{Partial Mutual Information Analysis of Financial Networks}
\shorttitle{Partial Mutual Information Analysis of Financial Networks} 

\author{P. Fiedor\inst{1}}
\shortauthor{P. Fiedor}

\institute{                    
  \inst{1} Cracow University of Economics - Rakowicka 27, 31-510 Krak{\'o}w, Poland\\
}

\abstract{
The econophysics approach to socio-economic systems is based on the assumption of their complexity. Such assumption inevitably lead to another assumption, namely that underlying interconnections within socio-economic systems, particularly financial markets, are nonlinear, which is shown to be true even in mainstream economic literature. Thus it is surprising to see that network analysis of financial markets is based on linear correlation and its derivatives. An analysis based on partial correlation is of particular interest as it leading to the vicinity of causality detection in time series analysis. In this paper we generalise the Planar Maximally Filtered Graphs and Partial Correlation Planar Graphs to incorporate nonlinearity using partial mutual information.}

\begin{document}

\maketitle

\section{Introduction}

The field of econophysics is perhaps best known for its treatment of financial markets as complex systems. In this treatment network theory plays an important role. Most commonly a correlation-based network structure is created, which quantifies the interrelations between financial instruments on a given market. Such analysis uncovers basic structure of the studied market but can also be useful for practical applications such as portfolio optimisation \cite{Markowitz:1952}. There have been numerous studies looking into stock markets in daily \cite{Mantegna:1999,Cizeau:2001,Forbes:2002,Podobnik:2008,Aste:2010,Kenett:2012meta} and intraday \cite{Bonanno:2001,Tumminello:2007,Munnix:2010} scales, as well as market indices \cite{Bonanno:2000,Maslov:2001,Drozdz:2001,Coelho:2007,Gilmore:2008,Eryigit:2009,Song:2011,Sandoval:2012} and foreign exchange markets \cite{McDonald:2005}. The well-corroborated results show that markets are structured according to sectors and sub-sectors of economic activities for stock markets and geographical locations for market indices and foreign exchange markets. Thus one can technically predict which sector does a financial instrument belong to, which is practically useless however. We see that there are limitations to this approach. One limitation is related to the above-mentioned limited usefulness. Much more interesting would be an analysis of causal relationships within the markets. This is done recently using either lead-lag effect (asymmetric correlations) or partial correlations. The second limitation is connected with the researchers' insistence on using strictly linear measure of closeness between studied elements of the market, even though it goes against the assumption of complexity of those systems \cite{Mantegna:2000,Rosser:2008} and the solid evidence of nonlinearity on financial markets with regards to stock returns \cite{Brock:1991,Qi:1999,McMillan:2001,Sornette:2002,Kim:2002}, market index returns \cite{Franses:1996,Wong:1995,Chen:1996,Wong:1997,Ammermann:2003}, and currency exchange rate changes \cite{Hsieh:1989,Brock:1991,Rose:1991,Brooks:1996,Wu:2003}. Recently we have been studying the financial networks using information-theoretic approach to account for nonlinearity \cite{Fiedor:2014a,Fiedor:2014lag}.

In this paper we are trying to address those issues together by using network analysis of financial markets based on partial mutual information. Partial mutual information is a generalisation of partial correlations, which is sensitive to nonlinear dependencies, to which Pearson's correlation and partial correlation are strictly not sensitive. Using partial mutual information allows us to either refine the classical structural analysis of the market by adding nonlinearity and controlling for mediating influence of third instruments. But partial mutual information may also be used to bring the analysis closer to market dynamics and causal relationships, similarly to the analysis performed with partial correlation in \cite{Kenett:2010,Kenett:2014}.

This paper is structured as follows: in Sec. 2 we present the proposed methodology. In Sec. 3 we show the results obtained for NYSE and briefly discuss them. In Sec. 4 we conclude our study and propose further research.

\section{Methods}

Our analysis is based on time series describing stock log returns, which is the standard way for analysing price movements. Thus data points are the log ratios between consecutive daily closing prices (this can be done at any other scale) \cite{Mantegna:2000}:
\begin{equation}
r_{t}=ln(p_{t}/p_{t-1}).
\end{equation}

For the purpose of estimating mutual information we need to discretize those data points. Thus we transform the data points into 4 distinct states. The states represent 4 quartiles, for the discussion of the significance and robustness of this step see \cite{Navet:2008,Fiedor:2014}.

On this basis we are estimating the partial mutual information (PMI), which is based on Shannon's entropy. For a discrete random variable $X$ with probabilities $p(x)$ of outcomes $\{x\}$, Shannon's entropy is defined as \cite{Shannon:1948}:
\begin{equation}
H(X)=-\sum_{x}p(x)\ln{p(x)}
\end{equation}

Mutual information (MI) of two random variables $X$ and $Y$ is given by:
\begin{equation}
I(X,Y)=H(X)+H(Y)-H(X,Y),
\end{equation}
where $H(X,Y)$ is obtained from the joint distribution of $(X,Y)$. MI is symmetric, and:
\begin{equation}
0\leq{}I(X,Y)\leq{}\min{\{H(X),H(Y)\}}.
\end{equation}
Mutual information measures information shared between two variables, thus both linear and nonlinear relationships between studied financial instruments are measured this way.

Then partial mutual information $I(X,Y|Z)$ denotes the part of mutual information $I(X,Y)$ that is not in $Z$ and is defined as:
\begin{equation}
I(X,Y|Z)=H(X,Z)+H(Y,Z)-H(Z)-H(X,Y,Z).
\end{equation}
PMI is symmetric so that $I(X,Y|Z)=I(Y,X|Z)$ and $0\leq{}I(X,Y|Z)$. MI and PMI are only equal to $0$ when $X$ and $Y$ are strictly independent.

To estimate PMI we need an estimator of Shannon's entropy. There is an abundance of estimators \cite{Paninski:2003,Stevens:1995,Strong:1998,Reinagel:2000,Nemenman:2004,Warland:1997}, in this study we use the Schurmann-Grassberger estimate of the entropy of a Dirichlet probability distribution, which is thought to be the best choice outside very specific conditions (particularly small samples) \cite{Bonachela:2008}. The Schurmann-Grassberger estimator is a Bayesian parametric procedure which assumes samples distributed following a Dirichlet distribution:
\begin{equation}
\begin{split}
&\hat{H}(X)=\frac{1}{m+|\chi|N} \\
&\sum_{x\in\chi}{(\#(x)+N)(\psi{}(m+|\chi|N+1)-\psi(\#(x)+N+1))},
\end{split}
\end{equation}
where $\#(x)$ is the number of data points having value $x$, $|\chi|$ is the number of bins from the discretization step, $m$ is the sample size, and $\psi(z)=\mathrm{d}\ln{\Gamma(z)}/\mathrm{d}z$ is the digamma function. The Schurmann-Grassberger estimator assumes $N=1/|\chi|$ as the prior \cite{Schurmann:1996}.

For comparison we have also used Pearson's correlation and partial correlation, and networks based on it, as defined in \cite{Kenett:2014}.

We now turn to the networks we are creating based on partial mutual information. First we refine the structural view of the market offered by the networks based on correlation. It is refined first by swapping correlation with mutual information, which adds nonlinearity. We may further refine this by removing the mediated parts of the interrelations between financial instruments by controlling for a third variable with partial mutual information. We take the minimum of the PMI calculated controlling for all other financial instruments, to show only the part of the mutual information not contained in other studied time series. Thus taking in mind the standard mutual information based metric \cite{Fiedor:2014a,Fiedor:2014lag} the distance used for the network topology is defined as:
\begin{equation}
d(X,Y)=H(X,Y)-\min_{Z\neq{}X,Y}{I(X,Y|Z)}
\end{equation}
On this basis we may create a network with topological restraints of our choosing, we calculate minimal spanning trees and planar maximally filtered graphs which we call PMIMST and PMIPMFG. These trees and planar graphs are created based on a list of $d(X,Y)$ sorted in increasing order. By starting from the first entry of the list, we add a corresponding link if and only if the resulting network is still a tree or a forest (PMIMST) or is still planar, i.e. it can be drawn on the surface of a sphere without link crossing (PMIPMFG).

Second we may refine the Partial Correlation Planar Graph as defined in \cite{Kenett:2010} by directly swapping partial correlation with partial mutual information. For this purpose we need a measure of MI influence or influence of an element $Z$ on the pair of elements $X$ and $Y$. This quantity is large only when a significant fraction of the PM $I(X,Y)$ can be explained in terms of $Z$. This measure is defined as:
\begin{equation}
d(X,Y|Z)=I(X,Y)-I(X,Y|Z)
\end{equation}
We define the average influence $d(X|Z)$ of element $Z$ on the MIs between element $X$ and all the other elements in the system as:
\begin{equation}
d(X|Z)=<d(X,Y|Z)>_{Y\neq{}X,Z}
\end{equation}

In order to construct a planar graph based on PMI we list the $N(N-1)$ values of the average MI influence $d(X|Z)$ in decreasing order. The construction protocol of the network begins by considering an empty network with $N$ vertices. By starting from the first entry of the list we put a link between them if and only if the resulting network is still planar. Similarly we can create a related minimal spanning tree by only adding a link if and only if the resulting network is still a tree or a forest.

Here we note that instead of filtering the information by network topology as we are doing above we may also filter the information by using a threshold to find which links should be entered into the network. Finding an appropriate threshold is not trivial however, but one can test for statistical significance based on the fact that $I(X,Y|Z)$ when $X$ and $Y$ are independent conditioned on $Z$ follows a Gamma distribution with shape parameter $\kappa=|Z|(|X|-1)(|Y|-1)/2$ and scale parameter $\Theta=1/N$ \cite{Kugiumtzis:2013}.

We also note a few areas in which this methodology can be extended. First we note that one could add market (index) as a second variable which is being controlled in PMI, thus eliminating the effects of general market trends on the relationships between financial instruments. Further we note that another approach to extending the analysis may be taken using transfer entropy, which is a measure closely related to partial mutual information \cite{Schindler:2007}. Transfer entropy is a measure quantifying causal information transfer between systems evolving in time, based on appropriately conditioned transition probabilities, thus it uses time lags. Time lagged causal analysis can also be performed using correlation \cite{Curme:2014} or mutual information \cite{Fiedor:2014lag} based method for filtering similarity measures into a network of statistically-validated directed links.

\section{Results}

To find the networks presented above we have taken log returns for 91 securities out of 100 which constitute the NYSE 100, excluding those with incomplete data. These log returns are based on daily closing prices. The data has been downloaded from Google Finance database available at http://www.google.com/finance/ and was up to date as of the 11th of November 2013, going 10 years back. The data is transformed in the standard way for analysing price movements, that is so that the data points are the log ratios between consecutive daily closing prices, as defined above, and those data points are, for the purpose of estimating mutual information, discretized into 4 distinct states.

Since there is no theory of financial markets we don't have any frame of reference to definitely state whether moving from correlation to mutual information and partial mutual information is bringing the analysis forward. We may only compare these two approaches and note how different the results are, and based on this we can risk an educated guess on the quality and usefulness of this method based on the fact that Shannon's mutual information is a more general measure than Pearson's correlation. We need to remember that while changing the particular methodology for creating networks of financial markets is changing the results, these changes will not be dramatic if the method is well-defined, thus graphically we won't see a completely different network. In other words the stocks should be clustered by economic sectors in all cases. But even small changes in these measurements may be of importance, particularly if this method is to be used in practice, especially automated analysis or trading. Then there is no reason to use correlation instead of mutual information, according to this study. This is particularly important for intraday price changes, where nonlinearity is significantly more prevalent \cite{Fiedor:2014}, which is important with the rising popularity of intraday and automated algorithmic trading.

Thus to analyse the differences we have created 6 similarity measures for the networks, and for each of them created a tree and a planar graph, creating 12 networks together. The distances used are as follows:
\begin{enumerate}
\item $d(X,Y)=\sqrt{2(1-\rho_{X,Y})}$;
\item $d(X,Y)=H(X,Y)-I(X,Y)$;
\item $d(X,Y)=\sqrt{2(1-\min_{Z\neq{}X,Y}{\rho_{X,Y})}}$;
\item $d(X,Y)=H(X,Y)-\min_{Z\neq{}X,Y}{I(X,Y|Z)}$;
\item $d(X|Z)=<d(X,Y|Z)>_{Y\neq{}X,Z}$ based on $\rho$;
\item $d(X|Z)=<d(X,Y|Z)>_{Y\neq{}X,Z}$ based on $I$.
\end{enumerate}

\begin{figure}
\onefigure[width=0.5\textwidth]{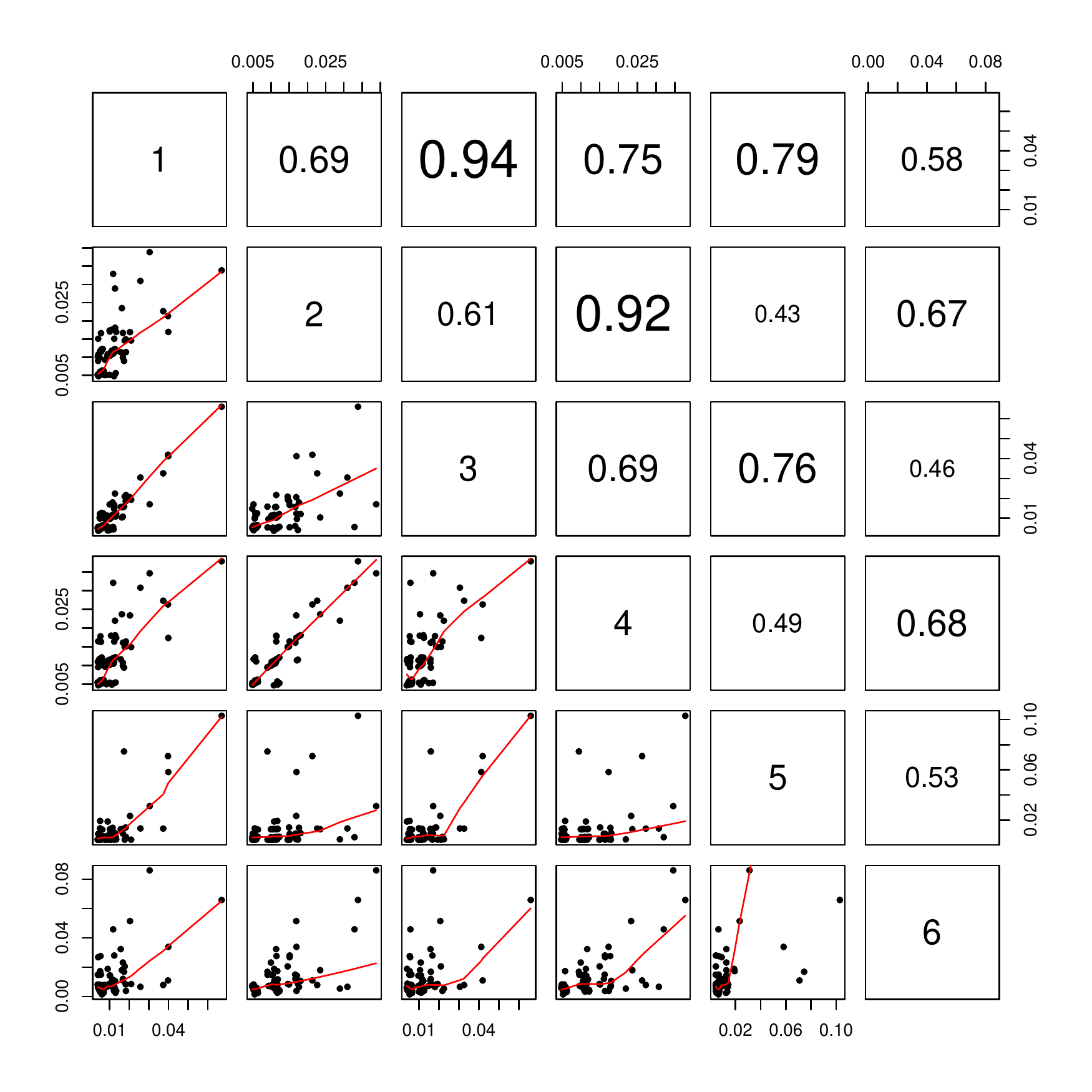}
\caption{MC correlations for trees}
\label{fig.5}
\end{figure}

\begin{figure}
\onefigure[width=0.5\textwidth]{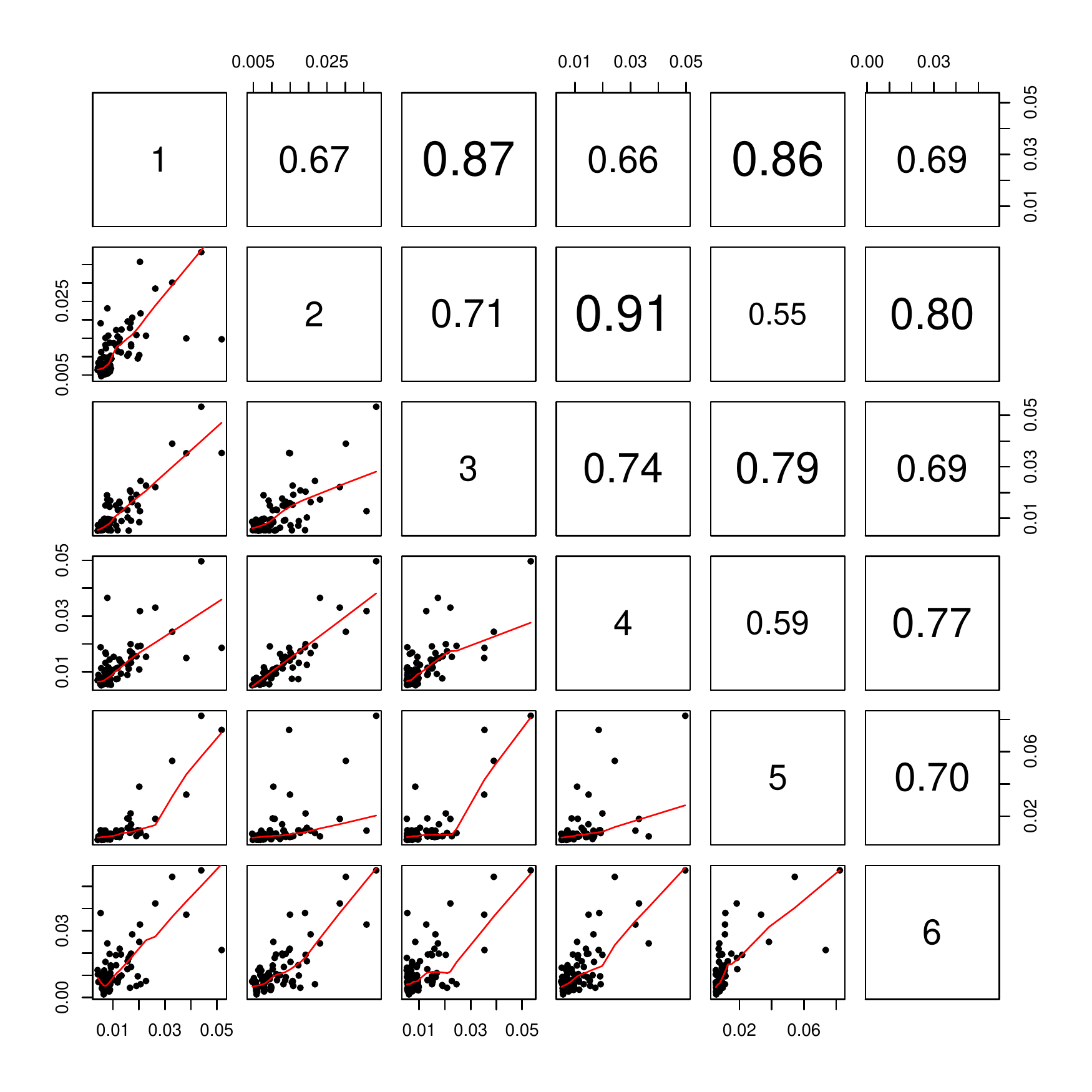}
\caption{MC correlations for planar graphs}
\label{fig.6}
\end{figure}

We can compare the networks on a node level, cluster level and network level. First at node level we calculate Markov centrality (for detailed description see \cite{Jeong:2001,White:2003,Brandes:2005}) for each node in all those networks and compare them between networks obtaining correlations presented in Figs. \ref{fig.5} (for trees) \& \ref{fig.6} (for planar graphs). We note that Markov centrality is basically a function of node degree, thus an analysis of node degree would be virtually identical. Similar analysis may be based also on node betweenness and other measures. Analyses for trees and graphs present similar data so we will comment collectively. We can see that the refinement of the standard methods by using minimal partial correlation/mutual information controlling for third variables is only changing the networks by about 10\% (noise from mediating relationships), while moving from correlation to mutual information changes the networks by about 30\%, hinting that around a third of relationships between financial instruments is nonlinear. Further we note that the dynamic approach is about 20-30\% different from the previous approaches, in this case the correlation and mutual information are giving much more varying results, thus indicating nonlinearity is more important in this dynamic assessment. In general we can conclude that the networks are significantly different (formal tests omitted due to space limitations) and thus the presented approach is valid, if the cluster level confirms this.

On cluster level we confirm that in all created networks the clusters have been aligned largely according to economic sectors, which means this important characteristic of the network approach to financial markets has been preserved. Preserving this information is important because it cannot be reproduced by simulating a virtual market \cite{Bonanno:2003}. This can be shown numerically as the ratio of arcs between stocks of the same sector to all nodes as presented in Table \ref{table.1}. As can be seen without any topological restraints the full market has 11.58\% links within sectors, but for our networks it's between 42\% and 67\%, thus we see that this important feature is preserved. In general we can also see that mutual information suits this task better, which further corroborates the usefulness of our approach.

On network level there is a limited possibility of investigation as most network-wide measures would be constrained by the common topological restraints we are using. Nonetheless we have calculated clustering coefficients (ratio of the number of triangles observed to the number of possible triangles in the network) for the planar graphs (there can't be any triangles in trees), which are presented in Table \ref{table.1} below. As can be seen mutual information produces significantly more clustering than correlation. Nonetheless partial correlation/mutual information analysis creates slightly less clustered networks, hinting that some clustering happens due to mediating noise, but also due to nonlinearity. The reference values in Table \ref{table.1} are calculated for a network with no topological restraints.

\begin{table}[htbp]
\caption{Network comparison}
\begin{tabular}{rrrr}
\hline
\multicolumn{1}{l}{Network} & \multicolumn{1}{l}{Tree ratio} & \multicolumn{1}{l}{Graph ratio} & \multicolumn{1}{l}{Clustering} \\ \hline
1 & 62.22\% & 49.06\% & 17.60\% \\
2 & 66.67\% & 55.81\% & 20.80\% \\
3 & 64.44\% & 51.69\% & 14.60\% \\
4 & 65.56\% & 54.68\% & 17.70\% \\
5 & 48.89\% & 42.70\% & 5.30\% \\
6 & 57.78\% & 47.19\% & 14.70\% \\
\multicolumn{1}{l}{Reference} & 11.28\% & 11.28\% & 50.00\% \\ \hline
\end{tabular}
\label{table.1}
\end{table}

\begin{figure*}
\onefigure[width=\textwidth]{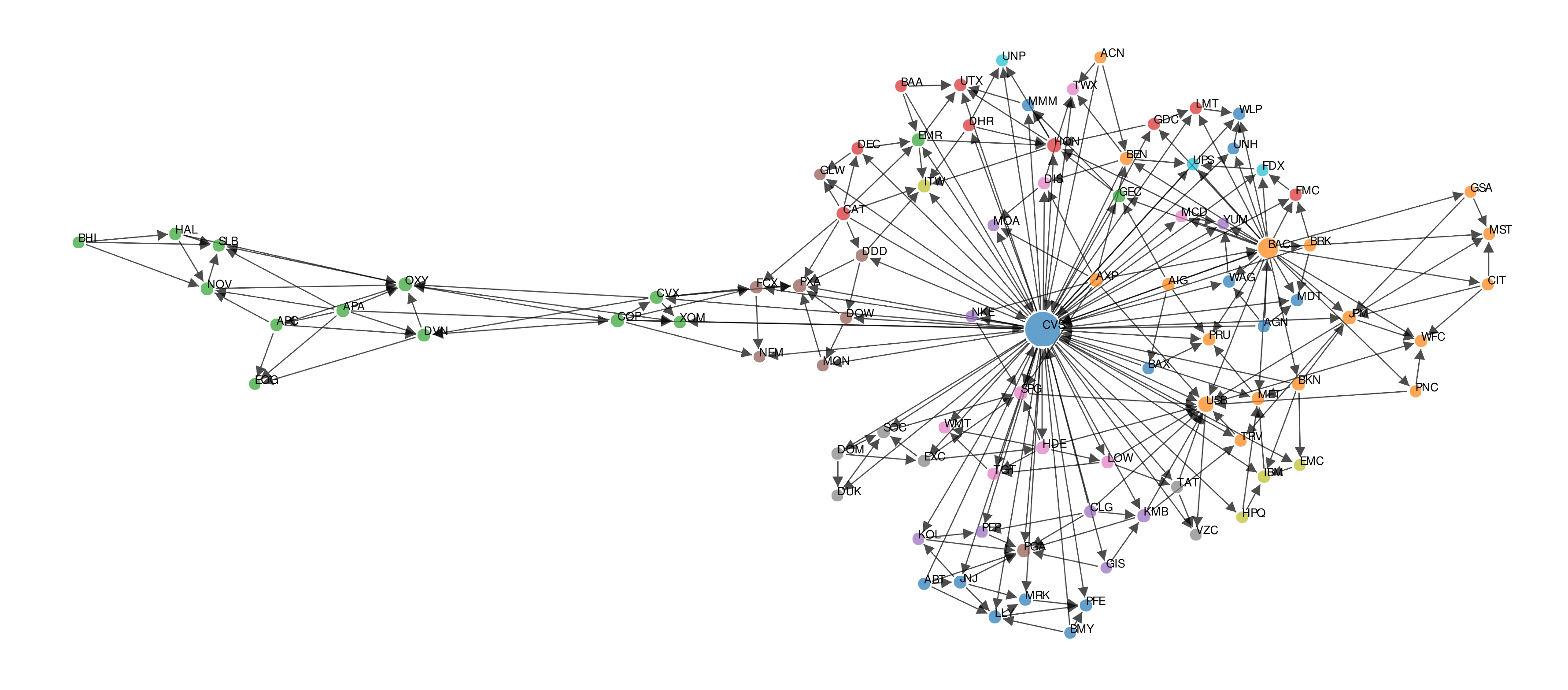}
\caption{Partial Mutual Information PMFG}
\label{fig.2}
\end{figure*}

\begin{figure*}
\onefigure[width=\textwidth]{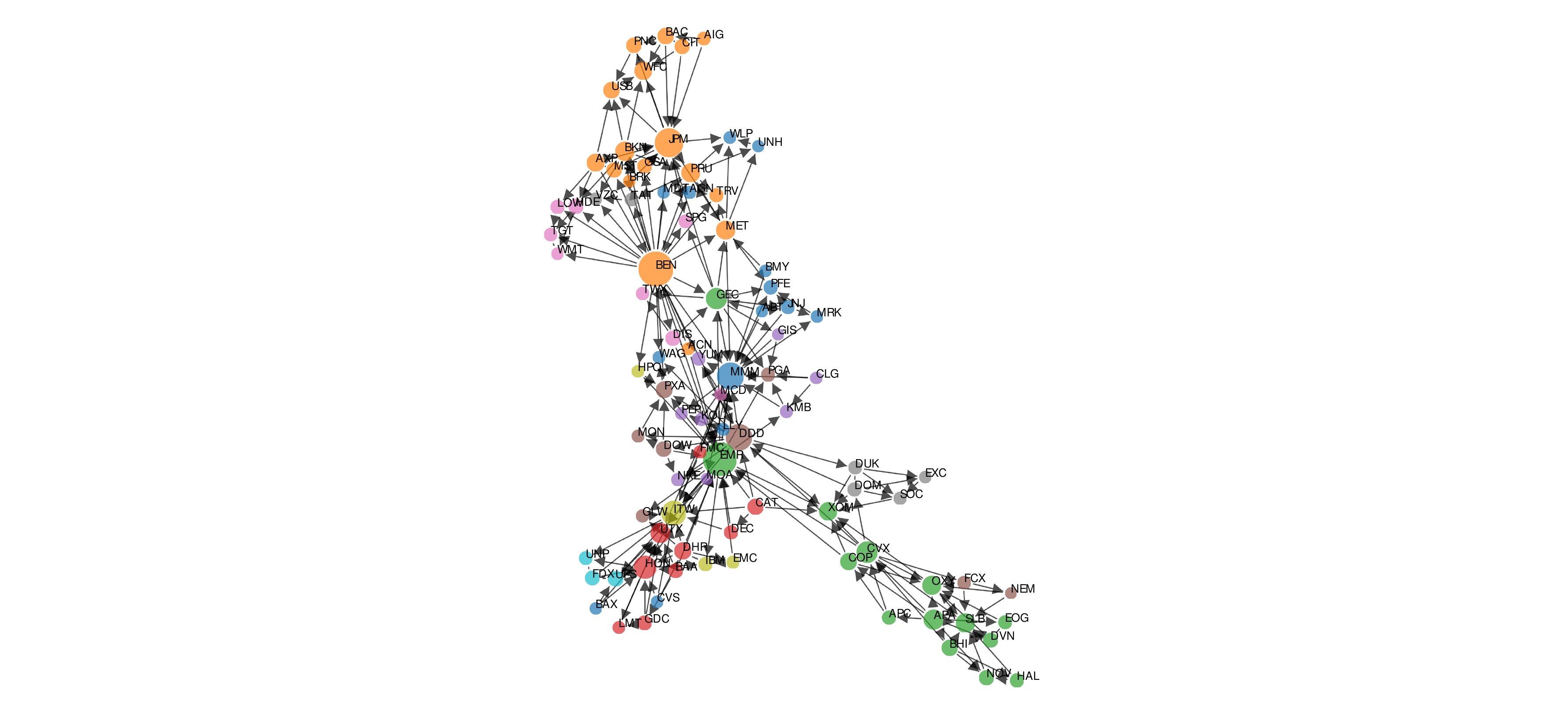}
\caption{Partial Mutual Information Planar Graph}
\label{fig.4}
\end{figure*}

Strictly as an example we have provided network number 4 on Fig. \ref{fig.2} and 6 on Fig. \ref{fig.4}. The node size is based on its centrality in the network.

\section{Conclusions}

We have analysed a method for producing networks of financial markets based on partial mutual information and how are those different from networks based on correlation, partial correlation and mutual information. The analysis leads us to believe that mutual information should be used in network analysis of financial markets as it provides different and likely more accurate networks, and that partial mutual information may be used to refine this analysis even further, particularly in controlling for mediating inference of third parties, thus making the network closer to real structure of the market. Further studies should be used based on other information theoretic measures such as transfer entropy, and also with controlling for more than one variable (particularly controlling for the whole market), as well as with other markets, and intraday price data.

\bibliographystyle{eplbib}
\bibliography{prace}

\begin{thebibliography}{10}
\expandafter\ifx\csname url\endcsname\relax\def\url#1{\texttt{#1}}\fi

\bibitem{Markowitz:1952}
\Name{Markowitz H.} \REVIEW{{J. Financ.}}{7}{1952}{77}.

\bibitem{Mantegna:1999}
\Name{Mantegna R.} \REVIEW{{Eur. Phys. J. B}}{11}{1999}{193}.

\bibitem{Cizeau:2001}
\Name{Cizeau P., Potters M. \and Bouchaud J.} \REVIEW{{Quant.
  Financ.}}{1}{2001}{217}.

\bibitem{Forbes:2002}
\Name{Forbes K. \and Rigobon R.} \REVIEW{{J. Financ.}}{57}{2002}{2223}.

\bibitem{Podobnik:2008}
\Name{Podobnik B. \and Stanley H.} \REVIEW{{Phys. Rev. Lett.}}{100}{2008}{}.

\bibitem{Aste:2010}
\Name{Aste T., Shaw W. \and Matteo T.~D.} \REVIEW{{New J.
  Phys.}}{12}{2010}{085009}.

\bibitem{Kenett:2012meta}
\Name{Kenett D., Preis T., Gur-Gershgoren G. \and Ben-Jacob E.}
  \REVIEW{{Europhys. Lett.}}{99}{2012}{38001}.

\bibitem{Bonanno:2001}
\Name{Bonanno G., Lillo F. \and Mantegna R.} \REVIEW{{Quant.
  Financ.}}{1}{2001}{96}.

\bibitem{Tumminello:2007}
\Name{Tumminello M., Matteo T.~D., Aste T. \and Mantegna R.} \REVIEW{{Eur.
  Phys. J. B}}{55}{2007}{209}.

\bibitem{Munnix:2010}
\Name{Munnix M., Schafer R. \and Guhr T.} \REVIEW{{Physica
  A}}{389}{2010}{4828}.

\bibitem{Bonanno:2000}
\Name{Bonanno G., Vandewalle N. \and Mantegna R.~N.} \REVIEW{{Phys. Rev.
  E}}{62}{2000}{7615}.

\bibitem{Maslov:2001}
\Name{Maslov S.} \REVIEW{{Physica A}}{301}{2001}{397}.

\bibitem{Drozdz:2001}
\Name{Drozdz S., Grummer F., Ruf F. \and Speth J.} \REVIEW{{Physica
  A}}{294}{2001}{226}.

\bibitem{Coelho:2007}
\Name{Coelho R., Gilmore C., B.Lucey, Richmond P. \and Hutzler S.}
  \REVIEW{{Physica A}}{376}{2007}{455}.

\bibitem{Gilmore:2008}
\Name{Gilmore C.~G., Lucey B.~M. \and Boscia M.} \REVIEW{{Physica
  A}}{387}{2008}{6319}.

\bibitem{Eryigit:2009}
\Name{Eryigit M. \and Eryigit R.} \REVIEW{{Physica A}}{388}{2009}{3551}.

\bibitem{Song:2011}
\Name{Song D.-M., Tumminello M., Zhou W.-X. \and Mantegna R.~N.} \REVIEW{{Phys.
  Rev. E}}{84}{2011}{026108}.

\bibitem{Sandoval:2012}
\Name{Sandoval L. \and Franca I.} \REVIEW{{Physica A}}{391}{2012}{187}.

\bibitem{McDonald:2005}
\Name{McDonald M., Suleman O., Williams S., Howison S. \and Johnson N.~F.}
  \REVIEW{{Phys. Rev. E}}{72}{2005}{046110}.

\bibitem{Mantegna:2000}
\Name{Mantegna R.~N. \and Stanley H.~E.} \Book{{Introduction to Econophysics:
  Correlations and Complexity in Finance}} ({Cambridge University Press}) 2000.

\bibitem{Rosser:2008}
\Name{Rosser B.} \REVIEW{Adv. Complex Syst.}{11}{2008}{745}.

\bibitem{Brock:1991}
\Name{Brock W.~A., Hsieh D.~A. \and LeBaron B.} \Book{{Nonlinear Dynamics,
  Chaos, and Instability. Statistical Theory and Economic Evidence.}} ({MIT
  Press}, Cambridge) 1991.

\bibitem{Qi:1999}
\Name{Qi M.} \REVIEW{{J. Bus. Econ. Stat.}}{17}{1999}{419}.

\bibitem{McMillan:2001}
\Name{McMillan D.} \REVIEW{{Int. Rev. Econ. Financ.}}{10}{2001}{353}.

\bibitem{Sornette:2002}
\Name{Sornette D. \and Andersen J.} \REVIEW{{Int. J. Mod. Phys.
  C}}{13}{2002}{171}.

\bibitem{Kim:2002}
\Name{Oh K. \and Kim K.} \REVIEW{{Expert Syst. Appl.}}{22}{2002}{249}.

\bibitem{Franses:1996}
\Name{Franses P.~H. \and Dijk D.~V.} \REVIEW{{J. Forecasting}}{15}{1996}{229}.

\bibitem{Wong:1995}
\Name{Abhyankar A., Copeland L. \and Wong W.} \REVIEW{{Econ.
  J.}}{105}{1995}{864}.

\bibitem{Chen:1996}
\Name{Chen P.} \REVIEW{{Stud. Nonlinear Dyn. E.}}{1}{1996}{}.

\bibitem{Wong:1997}
\Name{Abhyankar A., Copeland L. \and Wong W.} \REVIEW{{J. Bus. Econ.
  Stat.}}{15}{1997}{1}.

\bibitem{Ammermann:2003}
\Name{Ammermann P.~A. \and Patterson D.~M.} \REVIEW{{Pac. Bas. Financ.
  J.}}{11}{2003}{175}.

\bibitem{Hsieh:1989}
\Name{Hsieh D.} \REVIEW{{J. Bus.}}{62}{1989}{339}.

\bibitem{Rose:1991}
\Name{Meese R. \and Rose A.} \REVIEW{{Rev. Econ. Stud.}}{58}{1991}{603}.

\bibitem{Brooks:1996}
\Name{Brooks C.} \REVIEW{{Appl. Financ. Econ.}}{6}{1996}{307}.

\bibitem{Wu:2003}
\Name{Qi M. \and Wu Y.} \REVIEW{{J. Empir. Financ.}}{10}{2003}{623}.

\bibitem{Fiedor:2014a}
\Name{Fiedor P.} \Book{{Mutual Information Rate-Based Networks in Financial
  Markets}} arXiv:1401.2548 (2014).

\bibitem{Fiedor:2014lag}
\Name{Fiedor P.} \Book{{Information-theoretic approach to lead-lag effect on
  financial markets}} (2014).

\bibitem{Kenett:2010}
\Name{Kenett D., Tumminello M., Madi A., Gur-Gershgoren G., Mantegna R. \and
  Ben-Jacob E.} \REVIEW{{PloS one}}{5}{2010}{e15032}.

\bibitem{Kenett:2014}
\Name{Kenett D., Huang X., Vodenska I., Havlin S. \and Stanley H.}
  \Book{{Partial correlation analysis: Applications for financial markets}}
  arXiv:1402.1405 (2014).

\bibitem{Navet:2008}
\Name{Navet N. \and Chen S.-H.} \Book{{On Predictability and Profitability:
  Would GP Induced Trading Rules be Sensitive to the Observed Entropy of Time
  Series?}} in \Book{{Natural Computing in Computational Finance}}, edited by
  \Name{Brabazon T. \and O'Neill M.} Vol. 100 of \emph{Studies in Computational
  Intelligence} (Springer) 2008.

\bibitem{Fiedor:2014}
\Name{Fiedor P.} \Book{{Frequency Effects on Predictability of Stock Returns}}
  presented at \Book{{Proceedings of the IEEE Computational Intelligence for
  Financial Engineering \& Economics 2014}} 2014.

\bibitem{Shannon:1948}
\Name{Shannon C.~E.} \REVIEW{Bell Syst. Tech. J.}{27}{1948}{379}.

\bibitem{Paninski:2003}
\Name{Paninski L.} \REVIEW{{Neural Comput.}}{15}{2003}{1191}.

\bibitem{Stevens:1995}
\Name{Stevens C. \and Zador A.} \REVIEW{{NIPS}}{}{1995}{75}.

\bibitem{Strong:1998}
\Name{Strong S., Koberle R., de~Ruyter~van Steveninck R. \and Bialek W.}
  \REVIEW{{Phys. Rev. Lett.}}{80}{1998}{197}.

\bibitem{Reinagel:2000}
\Name{Reinagel P.} \REVIEW{{Current Biology}}{10}{2000}{542}.

\bibitem{Nemenman:2004}
\Name{Nemenman W., Bialek W. \and de~Ruyter~van Steveninck R.} \REVIEW{{Phys.
  Rev. E}}{69}{2004}{056111}.

\bibitem{Warland:1997}
\Name{Warland D., Reinagel P. \and Meister M.} \REVIEW{{J.
  Neurophysiol.}}{78}{1997}{2336}.

\bibitem{Bonachela:2008}
\Name{Bonachela J., H H. \and Munoz M.} \REVIEW{{J. Phys. A: Math.
  Theor.}}{41}{2008}{202001}.

\bibitem{Schurmann:1996}
\Name{Schurmann T. \and Grassberger P.} \REVIEW{{Chaos}}{6}{1996}{414}.

\bibitem{Kugiumtzis:2013}
\Name{Kugiumtzis D.} \REVIEW{{Eur. Phys. J. Spec. Top.}}{222}{2013}{401}.

\bibitem{Schindler:2007}
\Name{Hlavackova-Schindler K., Palus M., Vejmelka M. \and Bhattacharya J.}
  \REVIEW{{Phys. Rep.}}{441}{2007}{1}.

\bibitem{Curme:2014}
\Name{Curme C., Tumminello M., Mantegna R., Stanley H. \and Kenett D.}
  \Book{{Emergence of statistically validated financial intraday lead-lag
  relationships}} (2014).

\bibitem{Jeong:2001}
\Name{Jeong H., Mason S.~P., Barabasi A.~L. \and Oltvai Z.~N.}
  \REVIEW{Nature}{411}{2001}{}.

\bibitem{White:2003}
\Name{White S. \and Smyth P.} \Book{{Algorithms for estimating relative
  importance in networks}} in proc. of \Book{{Proceedings of the ninth ACM
  SIGKDD international conference on Knowledge discovery and data mining}}
  ({ACM}) 2003 pp. 266--275.

\bibitem{Brandes:2005}
\Name{Brandes U. \and Erlebach T.} \Book{{Network Analysis : Methodological
  Foundations}} {Lecture Notes in Computer Science} (Springer) 2005.

\bibitem{Bonanno:2003}
\Name{Bonanno G., Caldarelli G., Lillo F. \and Mantegna R.} \REVIEW{{Phys. Rev.
  E}}{68}{2003}{046130}.

\end{thebibliography}
\end{document}